\documentclass[prb,twocolumn,preprintnumbers,amsmath,amssymb]{revtex4}
\usepackage{graphicx}
\usepackage{color}
\usepackage{multirow}
\usepackage[sort&compress]{natbib}

\newcommand{\eV}{\ensuremath{\,\mbox{eV}}}

\newcommand{\mV}{\ensuremath{\,\mbox{mV}}}

\newcommand{\nm}{\ensuremath{\,\mbox{nm}}}

\newcommand{\vc}[1]{\ensuremath{\vec{#1}}}

\newcommand{\Hop}{\ensuremath{\mathcal{H}}}
\newcommand{\Hmb}{\ensuremath{\mathcal{H}}}

\newcommand{\NAtoms}{\ensuremath{M}}

\begin{document}

\title{A many-electron tight binding method for the analysis of quantum dot systems}
\author{Erik Nielsen, Rajib Rahman, and Richard P.~Muller}
\affiliation{Sandia National Laboratories, Albuquerque, New Mexico 87185 USA}
\date{\today}

\begin{abstract}
We present a method which computes many-electron energies and eigenfunctions by a full configuration interaction which uses a basis of atomistic tight-binding wave functions.  This approach captures electron correlation as well as atomistic effects, and is well suited to solid state quantum dot systems containing few electrons, where valley physics and disorder contribute significantly to device behavior.  Results are reported for a two-electron silicon double quantum dot as an example.
\end{abstract}

\maketitle

\section{Introduction}

Semiconductor devices have traditionally been simulated and understood using semiclassical band theory.  An increasing number of today's nanoscale semiconductor devices, however, require more precise treatment of their quantum mechanical nature.  Particularly within the area of quantum computation, capturing quantum effects is crucial to understanding the behavior and operation of devices.

Simulation of quantum-mechanical semiconductor devices presents several challenges.  The many degrees of freedom present due to the semiconductor lattice and the interactions between electrons quickly lead to intractable problems.  Effective mass theory significantly reduces the number of degrees of freedom by approximating the effects of the lattice with several effective parameters, but as a result is unable to capture atomistic effects such as interface roughness, lattice miscuts, defects, alloy disorder, and valley splitting without introducing adjustable parameters.  

There is also a question of how approximately the electron-electron interactions can be treated in a quantum device.  In the many-electron devices used in classical computing the interactions among electrons can be well approximated using Fermi liquid theory or by adding effective potential terms, and in any case the precise wave function of the many-electron system is not nearly as important as the electron density.  Devices proposed for quantum computation, however, often rely on manipulation of a wave function of few electrons and accurate modeling of electron correlation is essential.

There are well-established methods to separately satisfy the demands to capture atomistic effects and electron-electron interactions. Tight binding methods \cite{Slater_Koster, Harrison}, which include orbitals for each atom of the lattice, are well-suited for modeling atomistic effects in crystalline materials. Although this method scales well in speed and memory with the device volume enabling several million atom systems to be solved rapidly \cite{Klimeck2}, it is essentially a single electron theory. Many-electron effects can only be captured in an approximate way by mean-field methods such as charge self-consistent iterations with the Poisson equation \cite{Rahman_CE} or a reduced basis Hartree Fock method \cite{Pavel}. Exact many-body exchange-correlations and wave functions are difficult to capture under this framework, especially for systems with a large number of atoms.

The configuration interaction (CI) method \cite{Szabo} is potentially one of the most precise ways of capturing electron-electron interactions.  In it, the full quantum Hamiltonian is exactly diagonalized in a basis of multi-electron states.  Since its computational requirements grow rapidly with particle number, the CI method is limited to systems with small numbers of electrons.   In this paper, we describe a method which integrates empirical tight binding and full configuration interaction methods.  We also present an algorithm which achieves a substantial speed up in the evaluation of the Coulomb and Exchange integrals, which is the principle bottleneck of past attempts to integrate the two methods \cite{Whaley, Lee_JPL, Lee_JPL_2}. The result is a tool which satisfies both of the issues discussed above simultaneously.  In particular, it is capable of accurately modeling few-electron quantum devices involving several million atoms while including atomistic effects.  Prior work has used tight-binding configuration interation methods for smaller systems such as molecules and nanocrystals,\cite{PariserParr_earlyMolecTBCI_1953,Martin_TightBindingCI_1994,Takagahara_TBCI_Nanocrystals_1996,Hill_nanocrysta1_TBCI_1996,Reboredo_smallSiDot_TBCI_1999,MOPAC2009} but employs different algorithms which are less well suited to the large number of atoms that must be considered quantum dot devices.

As this method is motivated by the need to simulate quantum dot systems to be used as quantum bits (qubits), we study as an example a silicon double quantum dot (DQD) qubit which encodes quantum information in the lowest energy two-electron singlet and triplet states \cite{PettaScience_2005, TaylorDQDs_2007}.  Understanding and quantitatively estimating effects arising from the multi-valley nature of silicon, as well as the disorder which occurs at heterostructure interfaces, is crucial to an accurate understanding of the device operation.  Furthermore, because quantum information is stored in two-electron states, precise accounting of the interaction between these electrons is required.  By design the simulation goals for this system align with our method's capabilities.  We note that more approximate methods, such as effective mass theory, may consider the effects which arise from the lattice but must then also include adjustable parameters which describe these effects \cite{Culcer1, Culcer2}.  In our method, such adjustable parameters are not necessary because atomistic quantities can be computed directly.

In the sections that follow, we describe the method and then present results for our example system.

\section{Method}

\subsection{Overview}
Our approach begins with an empirical tight binding solver, NEMO3D \cite{Klimeck1, Klimeck2} (Nanoelectronic Modeling Tool), which models a system as a collection of atoms, each with a specified number of orbitals, within the tight binding approximation.  This accounts for atomic properties of the material being simulated, such as valley coupling, defects, lattice strain, and interface roughness.  The single-particle wave functions and energies generated by NEMO3D are input to a configuration interaction method which diagonalizes the many-body Hamiltonian in the basis of Slater determinants constructed from NEMO3D wave functions. Some of the main features of the two methods are described below.

\subsubsection{Empirical tight binding method in NEMO3D}
In the empirical tight-binding method, the Hamiltonian of the lattice is described in a basis of atomic orbitals localized in each atom. As a result, the electronic wave function is expressed as a linear combination of these atomic orbitals \cite{Harrison}. Following Slater and Koster \cite{Slater_Koster}, best results are obtained if the Hamiltonian matrix elements are optimized to fit critical features of the bulk band structure, such as band gaps, band minima locations, band symmetries, effective masses, and so on. Once a set of tight-binding Hamiltonian parameters are found for a host unit cell, they can used to compute the detailed electronic structure of any superlattice of the material along with an added external potential. This is a standard technique referred to as the empirical tight binding method.

In this work, we utilize the more complete 10 band sp$^3$d$^5$s* nearest neighbor model for silicon to represent the Hamiltonian \cite{Klimeck1}. The model parameters are optimized by an advanced genetic algorithm procedure \cite{Klimeck4}, and are well established in literature. The method, in general, is able to capture the detailed electronic structure of systems of different dimensions, such as bulk, quantum wells, wires and dots with respect to experiments \cite{Klimeck3, Shaikh, Rahman_prl, Rogge, Kharche}. The approach outlined in this paper is not specific to this band model, but can be applied generally to other band models and other materials. If the corresponding 20 band spin model is used, spin-orbit and Zeeman splittings can be captured directly from the eigen spectrum.

{\it{Bandstructure:}} Since the tight binding method is a full band-structure method, quantum states near all conduction or valence band valleys can be captured. Effects such as valley-splitting can be obtained without any additional adjustable parameters. 

{\it{Strain}:} If the device has at least two materials of different lattice constants, the relaxed structure is obtained by minimizing the strain energy by the valence force field method of Keating \cite{Keating}. This captures the effects of inhomogeneous strain distribution near hetero-structure interfaces, where quantum dot wave functions usually reside. Alternately, homogeneous strain can be simulated by setting the lattice constants to the relaxed values given by analytic expressions for various alloys \cite{Ioffe}. The strain induced modifications to the Hamiltonian matrix elements have been worked out in \cite{Harrison, Boykin_strain}.

{\it{Alloys}:} Alloyed materials such as Si$_{1-x}$Ge$_x$ can be treated both atomistically \cite{Boykin_sige} or from a virtual crystal approximation. An atomistic treatment allows the investigation of realistic systems relevant to experiments where disorder causes sample-to-sample variations in the measured electronic properties.  

{\it{External fields:}} Electrostatic potential either generated analytically or in TCAD (Technology Computer Aided Design) tools can be interpolated on the atomistic lattice, and can be included in the Hamiltonian. Hence, it is possible to study electronic states as a function of gate bias, which is the basis of all quantum nanoelectronics. Magnetic fields can be included both in symmetric or asymmetric gauges depending on the system. 

{\it{Geometry and interfaces:}} Complex device geometries can be described as a collection of simple shapes such as cuboids, cylinders, spheres, and so on. In general, realistic device geometries described in other tools can be imported, as long as the materials in the simulation domain are either crystalline or are described by a virtual crystal model. Hetero-structure interfaces are modeled naturally as the atom positions are defined. Hydrogen passivated interfaces have also been modeled \cite{Lee}. Moreover, miscut interfaces and surface roughness can also be defined atomistically, and used in the simulations \cite{Kharche}.

{\it{Solution:}} The full Hamiltonian is solved by a parallel Lanczos algorithm to obtain the desired number of eigenstates in a specified energy range of interest. If degenerate or nearly degenerate eigen states are required, a Block Lanczos algorithm is used with a block size chosen to meet the degree of resolution of the eigen states required \cite{Maxim}. 

{\it{Capability:}} NEMO3D is capable of solving large device volumes. Single electron wave functions of a few coupled quantum dots can be solved fairly quickly. The code has been demonstrated so far to solve 52 million atoms (101 nm$^3$) for electronic structure and over 100 million atoms for strain relaxation with very impressive scaling in memory and solution time \cite{Klimeck1}. 

\subsubsection{Configuration Interaction}

In the CI method, a basis of many electronic configurations represented by Slater Determinants \cite{Szabo} are constructed using the single electron wave functions. Each configuration corresponds to electrons occupying particular single electron energy levels. A Slater Determinant in which an electron has been promoted to an excited orbital from the ground state represents a singly excited configuration of the system. For an $N$ electron system, the highest excitation of the system consists of all $N$ electrons occupying a subset of the excited orbitals. If the basis set for the $N$ electron CI consists of all excitations, it is termed a full CI (FCI). In cases where a full CI is intractable, the number of excitations considered can be limited.

The $N$-electron many body Hamiltonian can be separated into single- and two-particle parts, $\Hop = \sum_{i=1}^N\Hop^{1P}_i + \sum_{i<j}\Hop^{2P}_{ij}$.  $\Hop^{1P}_i$ is a diagonal matrix in the basis of the single electron states $\psi_i$, with eigen energies $E_i$, which are obtained from NEMO3D.  The two-particle part is simply the Coulomb interaction, $\Hop^{2P}_{ij} = e^2/|\vc{r}_i-\vc{r}_j|$. This Hamiltonian is evaluated in the basis of the Slater Determinants to obtain the CI matrix. The technical details of an FCI method based on Gaussian orbitals has been described in our earlier work \cite{NielsenExchangePaper}.  The CI matrix is diagonalized using a Lanczos sparse matrix algorithm.  When the Hamiltonian commutes with total spin and/or the total of a component of the spin (\emph{e.g.}~total spin $z$-component), the CI matrix is first block diagonalized according to these symmetries and then each block is diagonalized separately. 

\subsection{Procedure}

The following is a step-by-step overview of the method.
\begin{enumerate}
\item System/device is specified within NEMO3D using regions of different materials, complex geometries, and external potential.  Domains for the computation of strain and electronic structure (not necessarily the same) are chosen.  
\item NEMO3D computes the $n$ lowest single particle energies $E_i$ and wave functions $\psi_i$ of the system, possibly performing lattice relaxation first to account for strain.
\item CI module discretizes each $\psi_i$ on a sub-atomic 3D rectangular grid by assuming analytic Slater-type orbital forms for the unknown basis functions used by NEMO3D.  This step is essentially performs a low-pass filter on the wave function. \label{stepSlaterApprox}
\item The Fourier transform is taken of every possible product of two discretized single particle wave functions, resulting in $n(n+1)/2$ ``pair products'' $\Pi_{ij}(\vc{r}) = \mathrm{FT}(\psi_i(\vc{r})\psi_j(\vc{r}))$ (discretized on the rectangular $k$-space lattice reciprocal to the spatial grid).  It is possible, at this stage, to only save portions of each pair product function with large magnitude (i.e. large spectral weight, determined by a cutoff), just as sparse matrices store only nonzero elements.  This sparse storage in the Fourier domain is similar to JPEG image compression, and can significantly reduce the storage requirements for the method. \label{stepPairProds}
\item 
The CI Hamiltonian is constructed using the single electron energies $E_i$ obtained from NEMO3D and the two-particle Coulomb interaction terms $\Hop^{2P}_{ij} = e^2/|\vc{r}_i-\vc{r}_j|$ evaluated using the pre-computed pair products as described in the detailed discussion below. \label{stepHam}.  
\item CI module diagonalizes the Hamiltonian matrix in the basis of Slater determinants of the $\psi_i$, and outputs the many-electron energies and wave functions.
\end{enumerate}

\subsection{Approximations}

Several approximations/assumptions are used in this method.  The first is the choice of the atomic orbitals for the tight-binding wave functions.  As we discussed earlier, in the empirical tight-binding method, the Hamiltonian matrix elements comprising on-site energies and nearest-neighbor interactions are optimized numerically to fit the bulk band structure of the host material. The symmetries of these orbitals determine which terms in the Hamiltonian are non-vanishing. No explicit forms of these orbitals are assumed. 

However, in order to evaluate the Coulomb and Exchange integrals within the CI, the spatial forms of these orbitals need to be specified. There is no other way to do this than to arbitrarily choose such a set of orbitals, as long as they adhere to the correct orbital symmetries and represent the atomic orbitals of the host to a certain degree. These basis sets cannot be related to the optimized tight-binding matrix elements in any way. The choice of the basis set can only be justified by comparing the many electron levels with those computed with other basis sets, and studying the sensitivity of the results on the choice. Benchmarking with experimental data can be another justification for the choice.

A prior work had developed a real space CI method based on tight-binding wave functions. Since the Coulomb and Exchange integrals were evaluated in real space rather than in momentum space, as we do here, the method is limited to dots of only 3-5 nm diameter. However, the work investigated the sensitivity of the many-electron states for a number of basis sets, such as Slater Type Orbitals, orthogonal and non-orthogonal Gaussian Type Orbitals. The work also showed that as the dot radii increased, the Coulomb and Exchange integrals became less sensitive to the choice of the basis set. In fact, beyond dots of radii 1.5-2.0 nm, the integrals were found quite reliable and representative of the actual values. This can be understood from the form of the integrals. If the Coulomb interaction is between two points a large distance apart, then the atomic orbitals appear as point-like charges, and their shapes do not influence the computation. Fortunately, the quantum dot wave functions we are considering span about 30 nm in each of the lateral dimensions. For such systems, the choice of the atomic basis set can be completely arbitrary. We have chosen STOs as they can be obtained easily for each material from the Slater rules \cite{Slater_rules}, and are representative of the atomic scale screening of the materials. 


We choose the following Slater-type forms for the two lowest s-, three lowest p-, and five lowest d-type symmetry orbitals, since they are good approximations to the orbitals of an isolated atom.  Expressions for these orbitals for silicon are given in Table \ref{tableBasisFns}. 
The need to choose a set of basis functions will arise whenever an empirical tight binding and configuration interaction techniques are coupled, and in this sense, the choice of a certain atomic basis is a necessary assumption for this method.

\begin{table}[h]
\begin{center}
\begin{tabular}{|c|l|}
\hline
Orbital type & wave function \\
\hline
$s$ & $1.30171 r^2 e^{-1.38 r}$ \\
$p$ & $1.30171 r^2 e^{-1.38 r} Y_{1m}(\theta,\psi)$ \\
$d$ & $0.037268 r^2 e^{-0.5 r} Y_{2m}(\theta,\psi)$ \\
$s*$ & $0.00337 r^{2.7} e^{-0.39 r}$ \\
\hline
\end{tabular}
\caption{Basis functions for silicon, as given by the Slater rules,\cite{Slater_rules} where $Y_{lm}$ is a spherical harmonic and $r$ is assumed to be in units of Bohr radii.\cite{SeungwonNote}\label{tableBasisFns}}
\end{center}
\end{table}



A second approximation is the discretization of the tight-binding wave function, which is a continuous function (but dependent on the specific choice of atomic basis functions), on a discrete uniform mesh (grid).  The constant spacing of this mesh determines a high-frequency cutoff for the resulting discrete function.  In order to capture atom-scale detail in the wave function, this cutoff must be large enough to capture the highest frequencies with appreciable weight of the Slater-type basis functions defined above.  This roughly corresponds to there being many mesh points on the scale of the basis functions, so that the shape of each basis function is well approximated using the given mesh.  When this criteria is met, the discretization represents only a minor approximation, especially since the exact form of the continuous basis functions are not known.  The reason for this discretization is performance-motivated, as will be described below, and other ways of coupling TB and CI methods do not require it.

The remaining approximations are those intrinsic to any CI method: a subset of the single particle functions, in our case the $n$ with lowest energy, are used to form the many-electron basis states.  In the appendix we analyze the convergence of the CI results with $n$ in a double quantum dot device.  The number of many-electron basis states can be reduced by limiting the solution space to $n_{ex}$ excitations relative to a given reference state (see Ref.~\cite{Szabo} for details on CI methods).  $n_{ex}$ can be between 1 and the number of particles $N$, at which point the CI is termed a ``full CI''. In the example shown in section \ref{secExample} below, $n=N=2$ so the CI is a full CI. 

\bigskip

\subsection{Performance}
The CI portion of the method, whether full or truncated, scales in the usual way with the number of particles $N$ and size of the single particle basis $n$.  In the full-CI case, this scaling is exponential in $N$ and $n$.  This is due to the size of the Hamiltonian matrix, whose construction and partial diagonalization usually dominate the computation.  When the single particle basis functions are defined on atomistic length scales, however, there is another potential bottleneck in the calculation: the evaluation of the two-electron interaction term $e^2/|\vc{r}_1-\vc{r}_2|$.  In cases where there are many atomic sites but few electrons, the computational resources required to construct the matrix $\Hop^{2P}$ (cf. step \ref{stepHam} above) can dominate even the diagonalization of the CI matrix.  Low-electron quantum-dot devices are a prime example of systems which have few electrons in regions containing many atomic sites.

It is to reduce the cost of evaluating the elements of $\Hop^{2P}$ that we perform the discretization in step \ref{stepSlaterApprox} and compute Fourier transforms of ``pair products'', $\Pi_{ij}$, in step \ref{stepPairProds}.  Let us denote the number of atomic sites as $\NAtoms$.  Since the number of mesh points at which we evaluate each single-particle wave function is proportional to the number of atomic sites, and since only a constant number of neighboring sites are needed to evaluate a wave function at a point (due to the localized nature of the atomic basis), the cost of evaluating each wave function on the discrete mesh is proportional to $\NAtoms$.  Multiplying two wave functions together and taking the fast Fourier transform of the result requires time $O(\NAtoms\ln(\NAtoms))$, so that in total the time required to compute the $n(n+1)/2$ pair products is $O(n\NAtoms + n^2\NAtoms\ln(\NAtoms)) = O(n^2\NAtoms\ln(\NAtoms))$.  After all $\Pi_{ij}$ are computed, the matrix elements of $\Hop^{2P}$ are computed as



\begin{widetext}
\begin{eqnarray}
\langle \psi_a \psi_b | \frac{e^2}{|\vc{r}_1-\vc{r}_2|} | \psi_c \psi_d \rangle &=& \int \bar{\psi}_a(\vc{r_1})\bar{\psi}_b(\vc{r_2}) \frac{e^2}{|\vc{r}_1-\vc{r}_2|} \psi_c(\vc{r_1}) \psi_d(\vc{r_2}) d\vc{r}_1 d\vc{r}_2 \\ 
&=& \int P_{ac}(\vc{r_1}) \frac{e^2}{|\vc{r}_1-\vc{r}_2|} P_{bd}(\vc{r_2}) \\
&=& \int \left(\int \Pi_{ac}(\vc{k_1})e^{i\vc{k}_1\vc{r}_1}d\vc{k}_1\right) \left( \int \frac{4\pi}{k_3^2}e^{i\vc{k}_3(\vc{r}_1-\vc{r}_2)}d\vc{k}_3 \right) \left(\int \Pi_{bd}(\vc{k_2})e^{i\vc{k}_2\vc{r}_2}d\vc{k}_2\right) d\vc{r}_1 d\vc{r}_2 \\
&=& (2\pi)^6 \int \Pi_{ac}(-\vc{k}) \frac{4\pi}{k^2} \Pi_{bd}(\vc{k}) d\vc{k} \,.
\end{eqnarray}
\end{widetext}
In the final line, we have an expression for an element of the $\Hop^{2P}$ matrix, in the basis of products $\psi_i\psi_j$, as a single (3D) integral over reciprocal space, with functions $\Pi_{ac}$ and $\Pi_{bd}$ precomputed as described.  As there are the same number of reciprocal space points as those of the real space mesh, the time to perform each of the $O(n^4)$ integrals is proportional to $\NAtoms$, and thus the time to construct the $\Hop^{2P}$ matrix is $O(n^4\NAtoms)$.

This can be compared with the more standard approach whereby each element $\langle \psi_a \psi_b | \frac{e^2}{|\vc{r}_1-\vc{r}_2|} | \psi_c \psi_d \rangle$ is computed as a double integral (or double sum over products of basis function coefficients) which requires time $O(n^4\NAtoms^2)$ to construct the $\Hop^{2P}$ matrix.  This extra factor of $\NAtoms$, which can be of order $10^6-10^7$, makes this direct method much slower than the one we describe (note the $O(n^2\NAtoms\ln(\NAtoms))$ time required to pre-compute the $\Pi_{ij}$ is insignificant).

\section{Example: Two-electron double quantum dot\label{secExample}}

We now apply the method to compute the two-electron states of a silicon DQD, an important implementation of a solid-state qubit. It was shown that the lowest single-triplet (i.e. unpolarized triplet $T_0$) of a DQD form a two-level system ideal to encode quantum information \cite{PettaScience_2005, TaylorDQDs_2007}. Recent experiments \cite{PettaScience_2005} have demonstrated spin initialization, coherent manipulation, and readout of this two-level system in a GaAs DQD. Due to the intrinsic nuclear spin associated with the host atoms in GaAs, the spins of the dot bound electrons decohere faster, even on the scale of nanoseconds to microseconds as $t_{2^*}$ measurements suggests \cite{PettaScience_2005}. Since silicon has a very low concentration of isotopes with nuclear spin and can be enriched to obtain high purity silicon, quantum confined electrons in silicon can have very long spin coherence times - even of order seconds \cite{Lyon}. Silicon DQDs are therefore an ideal choice for long-lived qubits. However, silicon systems are more prone to disorder at the Si-SiO$_2$ interface arising from charge defects or surface roughness. In addition, the conduction band valley degeneracy adds an additional degree of complexity in the electronic structure \cite{Boykin_VS, Friesen_apl, Friesen_prb, Kharche, Koiller_VS, Srikant, Culcer1}. 

Our method is ideal to study a silicon DQD in the presence of such imperfections. We show 2-electron energy level structure which includes valley effects and agrees with effective mass calculations.  The valley splitting appears without introducing an explicit valley term in the Hamiltonian, and we see that this splitting varies as expected with the vertical electric field strength.     

Consider a two-electron double quantum dot given by the Hamiltonian
\begin{equation}
\Hmb = \sum_{i=1}^2 \frac{(\vc{p}_i-e\vc{A})^2}{2m} + V(\vc{r}_i) + g\mu_B\vc{S}_i \cdot \vc{B} + \sum_{i < j}\frac{e^2}{\kappa |\vc{r}_i-\vc{r}_j|} \,,
\label{eqCIHam}
\end{equation} 
where $\vc{r}_i$ and $\vc{p}_i$ are the position and momentum, respectively, of the $i^{\mathrm{th}}$ electron, $V$ is the electrostatic potential, $m$ is the (bare) electron mass, and $\kappa$ is the silicon dielectric constant.  A vector potential $\vec{A}$ determines the magnetic field $\vc{B} = \vc{\nabla} \times \vc{A}$, which we set to zero for the cases considered here.  The potential is the sum of the potential due to the atomic lattice and an external potential, $V=V_{latt} + V_{ext}$.  We idealize the DQD potential $V_{ext}$ as the minimum of two parabolic dots with a vertical electric field $F_z$,
\begin{equation}
V_{ext}(x,y) = a \left[ \min\left( (x-L)^2+\epsilon, (x+L)^2 \right) + y^2\right]  \\ + F_zz\,. \label{eqPot}
\end{equation}
The parameters $\epsilon$, $L$, and $a$, correspond to the bias, inter-dot distance, and a measure of the confining well potential or dot size, respectively. $V_{latt}$ is set by the material(s); in this case silicon with a Hydrogen-passivated surface.

Qubit manipulation in the DQD system relies on a voltage controlled exchange splitting between the singlet and the triplet state. For this reason, one dot is detuned relative to the other by a gate bias $\epsilon$ such that the electrons in the dots make an adiabatic transition from a (1,1) occupation to a (0,2) occupation. A (1,1) occupation refers to one electron in each dot, a system in which the overlap between the electronic wave functions at the tunnel barrier between the dots is small or negligible, resulting in a very small exchange energy. As the dots are swept into the (0,2) occupation, which refers to both electrons in the same dot, the system undergoes a transition to a high overlap system with increased exchange energy. As the detuning is performed adiabatically, regions of the exchange curve in between these two extremums are also accessible. 

While this control sequence is well-established in a GaAs DQD \cite{TaylorDQDs_2007}, a Si DQD has additional valley degrees of freedom the impact of which depends on atomic scale effects such as miscuts, surface roughness, alloy disorder, and applied fields. The electronic states in a Si DQD therefore appears quite different from those in the GaAs system. We now present results for the DQD system which include these effects to illustrate the capabilities of this method.   


\subsection{Valley splitting}
\begin{figure}
\begin{center}
\includegraphics[angle=270,width=3.5in]{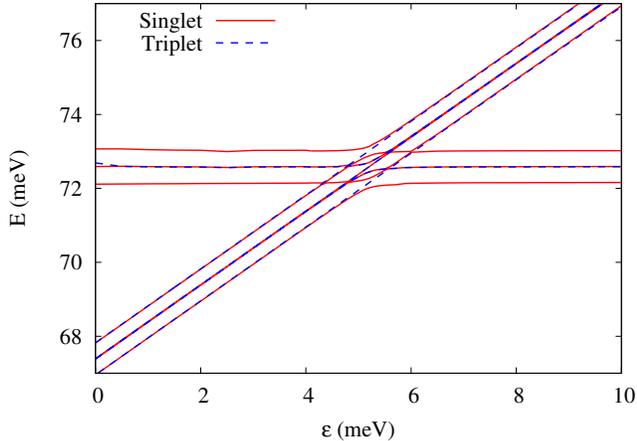}
\caption{Energy levels of a silicon (passivated-surface) DQD as a function of detuning parameter $\epsilon$.  $L=30\nm$, $a=0.0001 \eV/\nm^2$, and $F_z = 5\mV/\nm$. \label{figSmallValleySplitting}}
\end{center}
\end{figure}

The splitting in energy between different valleys in semiconductor arises from a material's atomic properties and crystal lattice symmetries.  In order to compute valley splitting from first principles, one must have an atomistic model.  Figure \ref{figSmallValleySplitting} shows the unpolarized ($S_z=0$) two-electron energy levels of a DQD parametrized by $L=20\nm$, $a=0.0001 \eV/\nm^2$, $F_z = 5\mV/\nm$.  We observe that the energy levels look like those of a DQD in a single valley material, but with a group of four states for each one state of the single valley case. This is due to the extra valley degree of freedom, as explained in Ref.~\cite{Culcer1}.  The separation between the central line and either the upper or lower line is the ``valley splitting'', and the spacings of the avoided crossings give a measure of the intra- and inter-valley coupling. We emphasize that these quantities are obtained without adjustable valley parameters, in contrast to effective mass methods.

\begin{figure}
\begin{center}
\includegraphics[angle=270,width=3.5in]{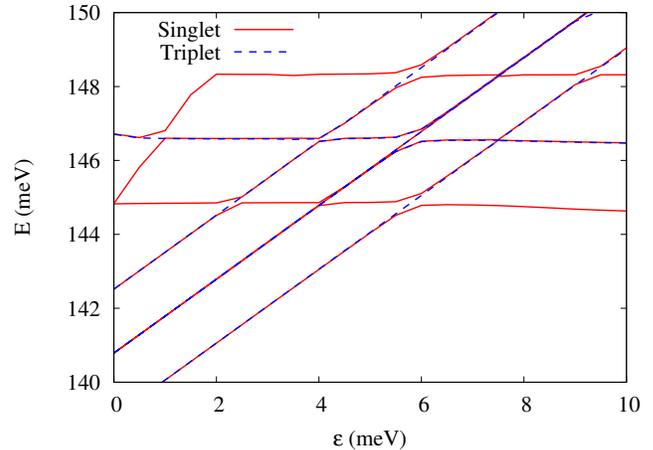}
\caption{Energy levels of a silicon (passivated-surface) DQD as a function of detuning parameter $\epsilon$.  $L=30\nm$, $a=0.0001 \eV/\nm^2$, and $F_z = 20\mV/\nm$. \label{figLargeValleySplitting}}
\end{center}
\end{figure}

Figure \ref{figLargeValleySplitting} shows how the energy levels of Fig.~\ref{figSmallValleySplitting} change upon increasing the vertical electric field $F_z$ to $20\mV/\nm$.  In particular, the valley splitting increases with larger $F_z$ as expected.


\subsection{Atomic perturbations}
Perturbations at atomistic length scales are nearly ubiquitous in solid state systems.  In this section we consider two types of atomistic perturbations, which we refer to as \emph{tilt} and \emph{roughness}.  By tilt, we mean that the silicon-oxide interface is not exactly parallel to a crystallographic plane. This occurs due to miscuts in the silicon wafer and the miscut angle is typically one half to a few degrees.  By roughness, we mean that the step edges at the silicon interface due to tilt are not straight but are located at a randomly varying positions. Since the QD electrons are confined at the interface between silicon and the barrier material, accurate models of interfaces are extremely important in electronic structure simulations. The TB method offers a natural way of dealing with tilt and surface roughness due to its atomistic nature.


Miscuts are always present in wafers because it is usually not possible to cut a wafer exactly along one of the crystallographic planes. Due to the miscut, the (001) surface of silicon reconstructs to form steps of monoatomic height and varying lengths. High precision scanning tunnelling microscope (STM) images have revealed the specific nature of these steps,\cite{roughness_paper1} and models of the reconstructed surfaces have been developed from equilibrium statistical mechanics based on the images.\cite{roughness_paper2} In this work, we have employed interface models of Refs.~\onlinecite{roughness_paper2, Kharche} to demonstrate the advantages of atomistic description of the interfaces for qubit simulations. 

\begin{figure}
\begin{center}
\begin{tabular}{cc}
\includegraphics[width=1.5in]{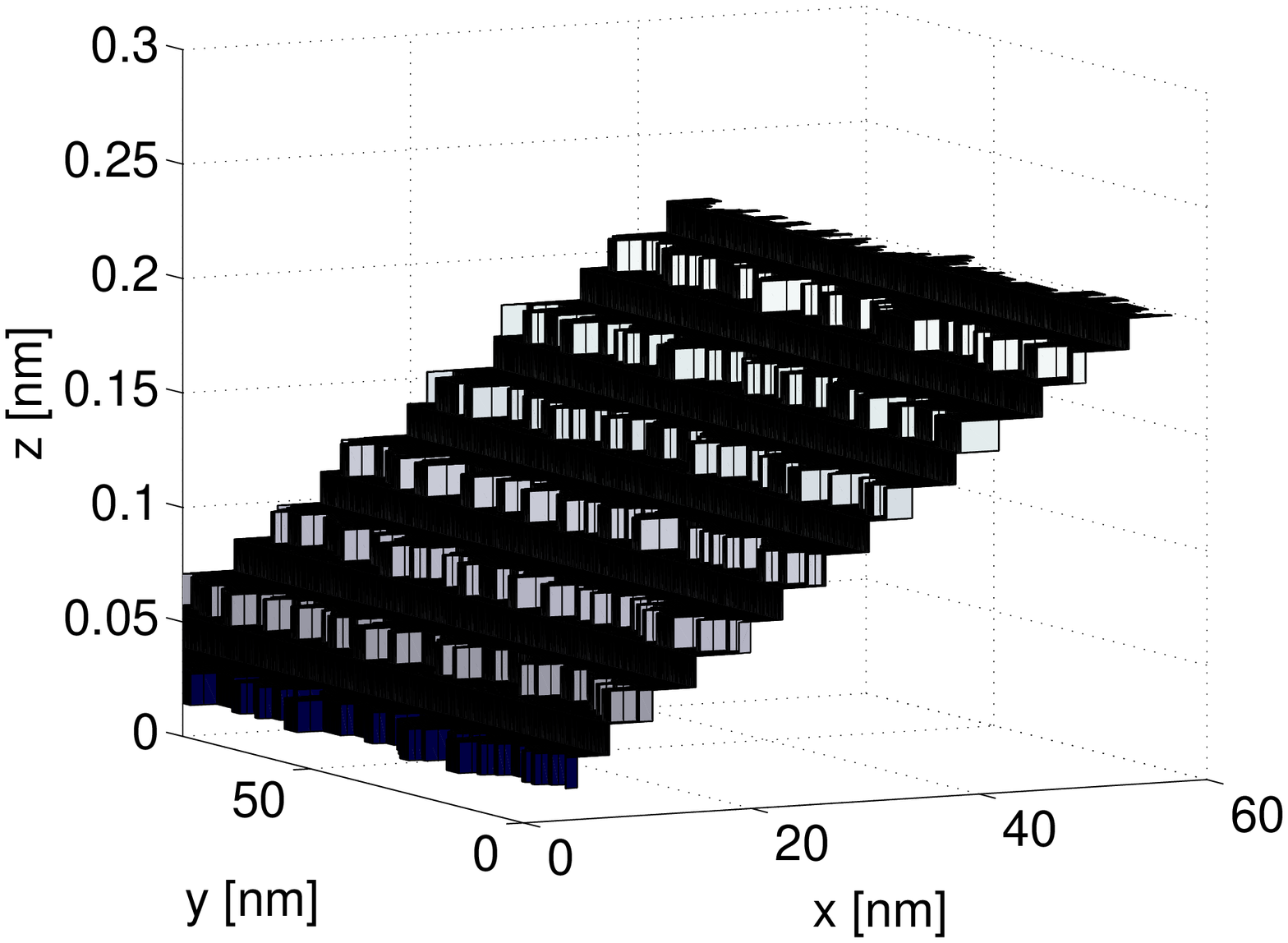} & \includegraphics[width=1.5in]{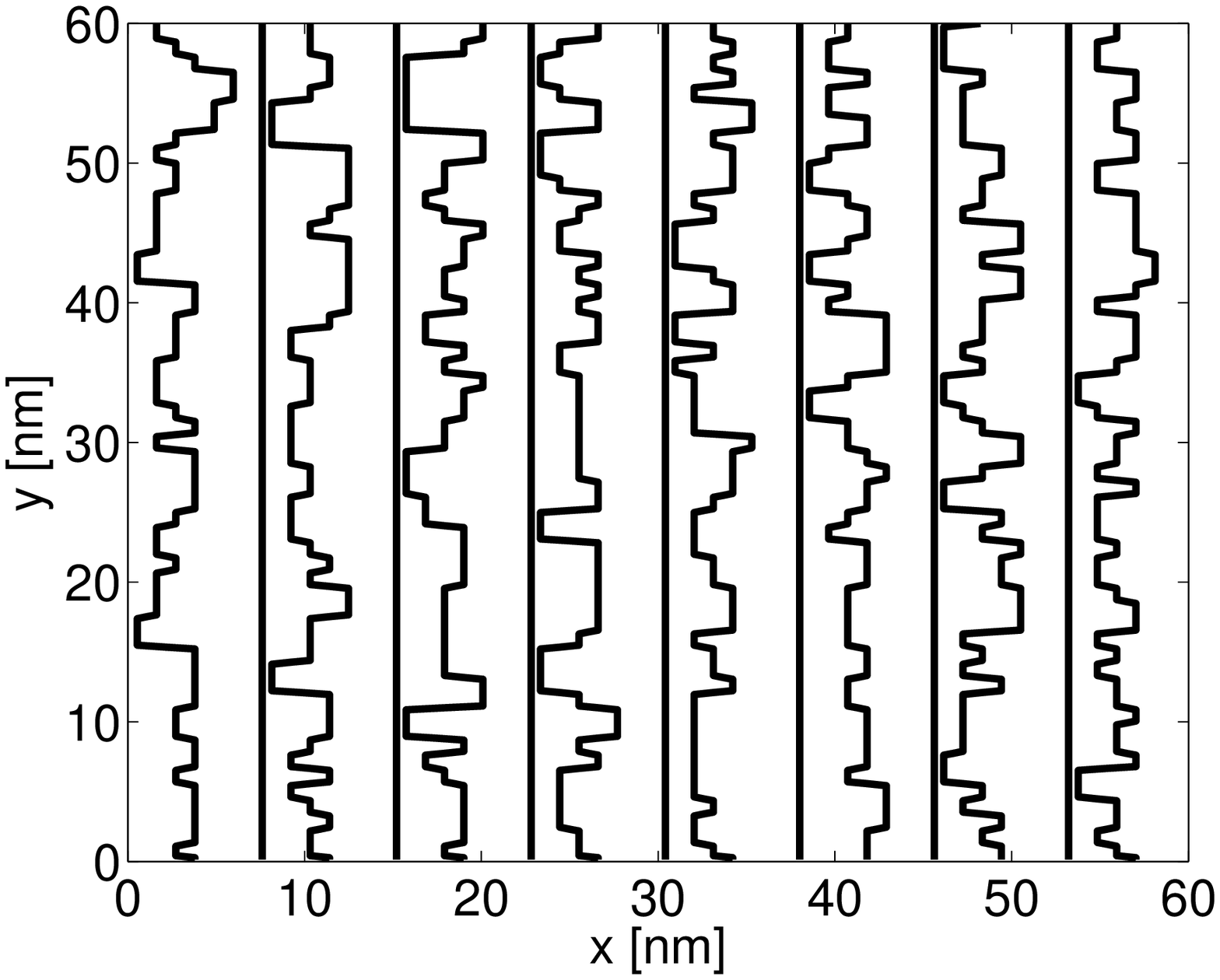} \\
(a) & (b)
\end{tabular}
\caption{A depiction of the interface roughness present at a silicon (001) surface, which consists of alternating straight and rough steps due to dimerized atoms (see text).  In (a) the real-space 3D profile of the surface is shown, and in (b) this view is projected onto the x-y plane.  Such realistic disorder can be modeled atomistically without adjustable parameters by the technique described.\label{figRoughnessPics}}
\end{center}
\end{figure}

In particular, the (001) surface of silicon reconstructs to form rows of dimerized atoms. As a result, two types of steps are noticed in the STM images, one set of steps running parallel and another perpendicular to the dimer rows. The parallel steps are straight, whereas the perpendicular steps have a high kink density, resulting in alternating straight and rough steps.\cite{roughness_paper1} A pictorial description of these steps is shown in Fig.~\ref{figRoughnessPics}, with a 3D plot in (a) and a 2D plot in (b). Based on this roughness profile, the interface is generated in TB by adding or removing sets of the surface atoms.\cite{Kharche} The addition of the CI technique now helps to understand the effect of these step disorder on the two-electron coupling essential in qubits.    

\begin{figure}
\begin{center}
\includegraphics[angle=270,width=3.5in]{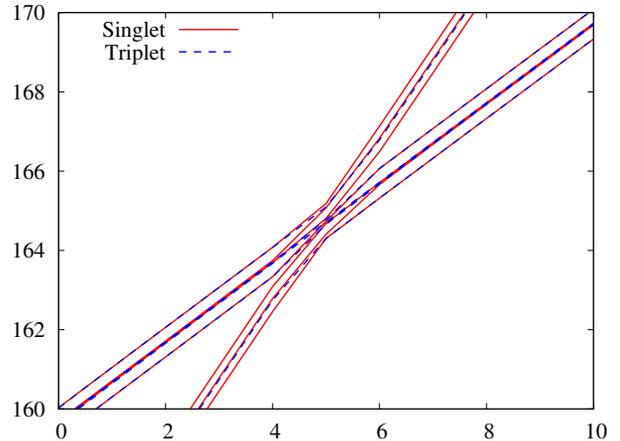}
\caption{Energy levels of a silicon (passivated-surface) DQD as a function of detuning parameter $\epsilon$.  The surface is tilted with respect to the crystallographic plane by 1 degree, and includes step roughness (the step edges due to the tilt are not straight).  These atomistic properties reduce the valley splitting considerably compared to the smooth surface case (Fig.~\ref{figLargeValleySplitting}). The slope of the lines relative to the smooth case is due to the vertical electric field not being perpendicular to the silicon surface and thereby contributing to the overall detuning of the DQDs. $L=30\nm$, $a=0.0001 \eV/\nm^2$, and $F_z = 20\mV/\nm$. \label{figEnergyInterfaceRoughness}}
\end{center}
\end{figure}

When the silicon interface is ideally tilted with no roughness, the valley splitting decreases relative to the smooth interface case due to interference between Bloch components of the wave function with different phases \cite{Friesen_apl, Kharche}.  This effect is clearly seen by comparing Figs.~\ref{figSmallValleySplitting} and \ref{figEnergyInterfaceRoughness}, the latter of which has a rough interface tilted by 1 degree.  We have chosen $L=20\nm$ in Fig.~\ref{figEnergyInterfaceRoughness} to better display the levels at low epsilon.  The case where $L=30\nm$, as in Fig.~\ref{figSmallValleySplitting} shows a similar reduction in valley splitting.  The addition of roughness results in washing away some of the interference effects produced by ideal steps due to the randomness. This increases valley splitting compared to the ideally tilted case (without roughness). However, the valley splitting is still less than that of a flat surface with no roughness. 

\section{Summary}
We have described a technique which couples tight binding and configuration interaction methods to make possible the accurate calculation of many-electron systems in the presence of atomistic effects such as valley splitting or disorder. To illustrate the method's capabilities we apply it to a silicon double quantum dot containing two electrons, where we find that varying the vertical electric field and adding interface tilt and roughness have the expected effects on the two electron states.

This work was supported by the Laboratory Directed Research and Development program at Sandia National Laboratories. Sandia National Laboratories is a multi-program laboratory managed and operated by Sandia Corporation, a wholly owned subsidiary of Lockheed Martin Corporation, for the U.S. Department of Energy's National Nuclear Security Administration under contract DE-AC04-94AL85000. RR acknowledges Gerhard Klimeck of Purdue University for the NEMO3D code, Seungwon Lee of NASA's Jet Propulsion Laboratory, Caltech, and Neerav Kharche of Rensselaer Polytechnic Institute for discussions.

\appendix

\section{ Convergence }
As explained in the text, the configuration interaction (CI) algorithm takes as input a fixed number of single particle levels, $n$ from which to generate Slater Determinant multi-electron basis functions.  It is important to check convergence with respect to $n$ in order to ensure that the basis being used by the CI is adequate at approximating the full quantum Hilbert space of wave functions.

\begin{figure}[h]
\begin{center}
\includegraphics[angle=270,width=3in]{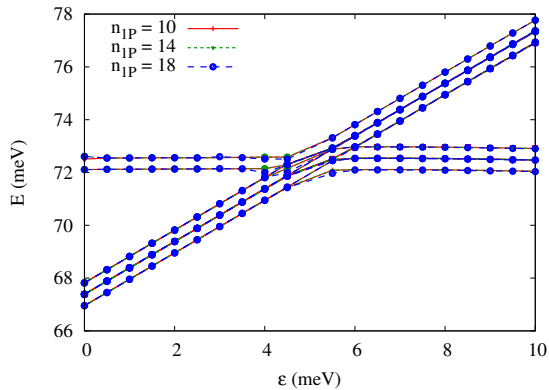}
\caption{Convergence of two-particle energies in a silicon DQD, given by the TB-CI method with respect to $n$, the number of NEMO3D wave functions used by the FCI.  DQD parameters $L=30\nm$, $a=0.0001 \eV/\nm^2$, and $F_z = 5 mV/\nm$. \label{figConvergence1}}
\end{center}
\end{figure}

\begin{figure}[h]
\begin{center}
\includegraphics[angle=270,width=3in]{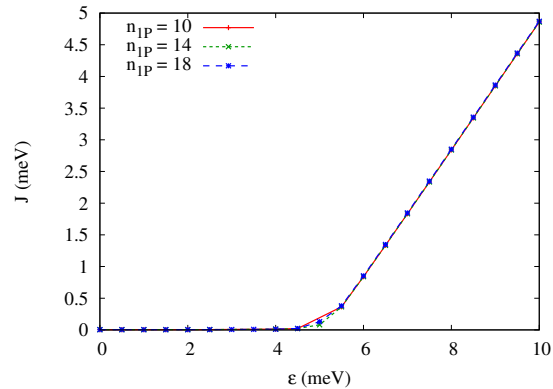}
\caption{Convergence of the exchange energy in a silicon DQD, given by the TB-CI method with respect to $n$, the number of NEMO3D wave functions used by the FCI.  DQD parameters $L=30\nm$, $a=0.0001 \eV/\nm^2$, and $F_z = 5 mV/\nm$. \label{figConvergence2}}
\end{center}
\end{figure}

  Figures \ref{figConvergence1} and  \ref{figConvergence2} show the energy levels and exchange (singlet-triplet) splitting, respectively, of a DQD for $n=10$, $14$, and $18$.  We find here and in general for quantum dots with radii from $15-25\nm$ that results which use $n=10$ are sufficiently converged.  The exchange energy in Fig.~\ref{figConvergence2} is taken to be the difference between the lowest singlet and unpolarized triplet states with the \emph{same} valley character.  (This means that at large values of $\epsilon$ we consider the lowest energy singlet and the lowest triplet of the group of states with positive slope.)  We expect when strong atomistic defect potentials are introduced to a quantum-dot system that a greater number of single particle levels will be required to obtain convergence.

\bibliography{nemoCI}

\begin{thebibliography}{43}
\bibitem[{Seu()}]{SeungwonNote}
\bibinfo{note}{We are indebted to Seungwon Lee of NASA’s Jet Propulsion
  Laboratory, Caltech, for providing us the computed numerical coefficients of
  this table}.
\end{thebibliography}

\end{document}